\documentclass[12pt]{article}
\usepackage{ssi}
\usepackage{times}
\usepackage{graphicx,here,rotate,epsf,epsfig}

\begin{document}

\title{Connections Between Big and Small}

\author{John Ellis\thanks{
\noindent
\copyright\ 2003 by John Ellis. }\\
Theory Division \\
CERN, CH-1211 Geneva 23, Switzerland \\[0.4cm]
}

%
%
%
 
\newcommand{\mycomm}[1]{\hfill\break{ \tt===$>$ \bf #1}\hfill\break}

\def\ga{\mathrel{\raise.3ex\hbox{$>$\kern-.75em\lower1ex\hbox{$\sim$}}}}
\def\la{\mathrel{\raise.3ex\hbox{$<$\kern-.75em\lower1ex\hbox{$\sim$}}}}
\def\gev{{\rm \, Ge\kern-0.125em V}}
\def\tev{{\rm \, Te\kern-0.125em V}}
\def\beq{\begin{equation}}
\def\eeq{\end{equation}}
\def\st{\scriptstyle}
\def\ss{\scriptscriptstyle}
\def\mb{m_{\widetilde B}}
\def\msf{m_{\tilde f}}
\def\mst{m_{\tilde t}}
\def\mf{m_{\ss{f}}}
\def\mpar{m_{\ss\|}^2}
\def\mpl{M_{\rm Pl}}
\def\mchi{m_{\chi}}
\def\ohsq{\Omega_{\chi} h^2}
\def\msn{m_{\tilde\nu}}
\def\m12{m_{1\!/2}}
\def\mstpl{m_{\tilde t_{\ss 1}}^2}
\def\mstpr{m_{\tilde t_{\ss 2}}^2}

\def\sm{Standard Model}

\def\ga{\mathrel{\raise.3ex\hbox{$>$\kern-.75em\lower1ex\hbox{$\sim$}}}}
\def\la{\mathrel{\raise.3ex\hbox{$<$\kern-.75em\lower1ex\hbox{$\sim$}}}}
\def\gyr{{\rm \, G\kern-0.125em yr}}
\def\gev{{\rm \, Ge\kern-0.125em V}}
\def\tev{{\rm \, Te\kern-0.125em V}}
\def\ss{\scriptscriptstyle}
\def\scs{\scriptstyle}
\def\mb{m_{\widetilde B}}
\def\mst{m_{\tilde\tau_R}}
\def\mstop{m_{\tilde t_1}}
\def\msl{m_{\tilde{\ell}_1}}
\def\stau{\tilde \tau}
\def\stop{\tilde t}
\def\sbot{\tilde b}
\def\mchi{m_{\tilde \chi}}
\def\mxi{m_{\tilde{\chi}_i^0}}
\def\mxj{m_{\tilde{\chi}_j^0}}
\def\mchari{m_{\tilde{\chi}_i^+}}
\def\mcharj{m_{\tilde{\chi}_j^+}}
\def\mgluino{m_{\tilde g}}
\def\msf{m_{\tilde f}}
\def\m12{m_{1\!/2}}
\def\mtb{\overline{m}_{\ss t}}
\def\mbb{\overline{m}_{\ss b}}
\def\mfb{\overline{m}_{\ss f}}
\def\mf{m_{\ss{f}}}
\def\gt{\gamma_t}
\def\gb{\gamma_b}
\def\gf{\gamma_f}
\def\thm{\theta_\mu}
\def\tha{\theta_A}
\def\thb{\theta_B}
\def\mgl{m_{\ss \tilde g}}
\def\cp{C\!P}
\def\ch{{\widetilde \chi}} 
\def\st{{\widetilde \tau}_{\scriptscriptstyle\rm 1}}
\def\sel{{\widetilde e}_{\scriptscriptstyle\rm R}}
\def\sl{{\widetilde \ell}_{\scriptscriptstyle\rm R}}
\def\msn{m_{\ch}}
\def\tsq{|{\cal T}|^2}
\def\tcm{\theta_{\rm\scriptscriptstyle CM}}
\def\half{{\textstyle{1\over2}}}
\def\neq{n_{\rm eq}}
\def\qeq{q_{\rm eq}}
\def\slash#1{\rlap{\hbox{$\mskip 1 mu /$}}#1}%
\def\mw{m_W}
\def\mz{m_Z}
\def\mhb{m_{H}}
\def\mhl{m_{h}}
\newcommand\f[1]{f_#1}
\def\nl{\hfill\nonumber\\&&}

\def\gappeq{\mathrel{\rlap {\raise.5ex\hbox{$>$}}
{\lower.5ex\hbox{$\sim$}}}}

\def\lappeq{\mathrel{\rlap{\raise.5ex\hbox{$<$}}
{\lower.5ex\hbox{$\sim$}}}}

\def\Toprel#1\over#2{\mathrel{\mathop{#2}\limits^{#1}}}
\def\FF{\Toprel{\hbox{$\scriptscriptstyle(-)$}}\over{$\nu$}}

\newcommand{\Zee}{$Z^0$}


\def\Yi{\eta^{\ast}_{11} \left( \frac{y_{i}}{2} g' Z_{\chi 1} + 
        g T_{3i} Z_{\chi 2} \right) + \eta^{\ast}_{12} 
        \frac{g m_{q_{i}} Z_{\chi 5-i}}{2 m_{W} B_{i}}}

\def\Xii{\eta^{\ast}_{11} 
        \frac{g m_{q_{i}}Z_{\chi 5-i}^{\ast}}{2 m_{W} B_{i}} - 
        \eta_{12}^{\ast} e_{i} g' Z_{\chi 1}^{\ast}}

\def\Wi{\eta_{21}^{\ast}
        \frac{g m_{q_{i}}Z_{\chi 5-i}^{\ast}}{2 m_{W} B_{i}} -
        \eta_{22}^{\ast} e_{i} g' Z_{\chi 1}^{\ast}}

\def\Vi{\eta_{22}^{\ast} \frac{g m_{q_{i}} Z_{\chi 5-i}}{2 m_{W} B_{i}}
        + \eta_{21}^{\ast}\left( \frac{y_{i}}{2} g' Z_{\chi 1}
        + g T_{3i} Z_{\chi 2} \right)}

\def\zthree{\delta_{1i} [g Z_{\chi 2} - g' Z_{\chi 1}]}

\def\zfour{\delta_{2i} [g Z_{\chi 2} - g' Z_{\chi 1}]}


\begin{center}
{\it Opening Lecture at the 31st SLAC Summer Institute, July 2003: 
PSN~L01}\\
\end{center}
\begin{center}
CERN-TH/2003-268 $\; \; \; \;$ {\tt astro-ph/0310911}
\end{center}

\maketitle 
\begin{abstract}%
\baselineskip 16pt 
Big-Bang cosmology and ideas for possible physics beyond the Standard
Model of particle physics are introduced. The density budget of the
Universe is audited, and the issues involved in calculating the baron
density from microphysics are mentioned, as is the role of cold dark
matter in the formation of cosmological structures. Candidates for cold
dark matter are introduced, with particular attention to the lightest
supersymmetric particle and metastable superheavy relics. Prospects for
detecting supersymmetric dark matter in non-accelerator experiments are
assessed, and the possible role of decays in generating ultra-high-energy
cosmic rays is discussed. More details of these and other astroparticle
topics are presented during the rest of this Summer Institute.
\end{abstract}

\section{Introduction}

My task in this opening lecture is to set the stage for the subsequent
lectures that develop in more detail the connections between particle
physics and cosmology. To do so, I first recall the essential aspects of
standard Big-Bang cosmology, emphasizing that the questions it raises
about the early history of the Universe can only be answered by particle
physics. The latter is described by its own Standard Model, which makes
successful quantitative predictions for accelerator experiments, but
leaves open many fundamental questions. These include the origin of
particle masses, the proliferation of different types of elementary
particles and the possible unification of all the particle interactions.  
In combination with accelerator experiments, astrophysics and cosmology
may cast important light on the solutions of these problems. According to
astrophysicists and cosmologists, most of the matter in the Universe has
never been seen, and cannot consist of ordinary matter~\cite{DM}. The
formation of structures in the Universe would be helped by presence of
massive weakly-interacting cold dark matter particles~\cite{SF}.  
Candidates for these include the lightest supersymmetric particle
(LSP)~\cite{EHNOS}, the axion~\cite{axion} and metastable superheavy
particles~\cite{Berez} whose decays might be responsible for 
ultra-high-energy cosmic
rays beyond the Greisen-Zatsepin-Kuzmin (GZK) cutoff~\cite{GZK}, if they 
exist.
These are just a few of the connections between the very big and the very
small that are developed by other lecturers at this Summer Institute.

\section{Big-Bang Cosmology}

According to standard Big-Bang cosmology~\cite{first3}, the entire visible 
Universe is expanding homogeneously and isotropically from a very dense and hot
initial state. The first direct piece of evidence for this was the 
discovery by Hubble that distant objects in the Universe are receding 
from each other at velocities proportional to their distances from each 
other:
\begin{equation}
v \; = \; H \cdot d,
\label{Hubblelaw}
\end{equation}
where $v$ is the recession velocity, $d$ is the distance and the Hubble 
constant $H \equiv h \cdot 100$~km/s/Mpc where $h \simeq 0.7$~\cite{HST}. 
Observations of the Universe suggest that it is indeed very homogeneous 
and isotropic on large scales $\gappeq 1000$~Mpc~\cite{2df}. 

\begin{figure}[htb]
\centering
\includegraphics[angle=90,width=0.8\textwidth]{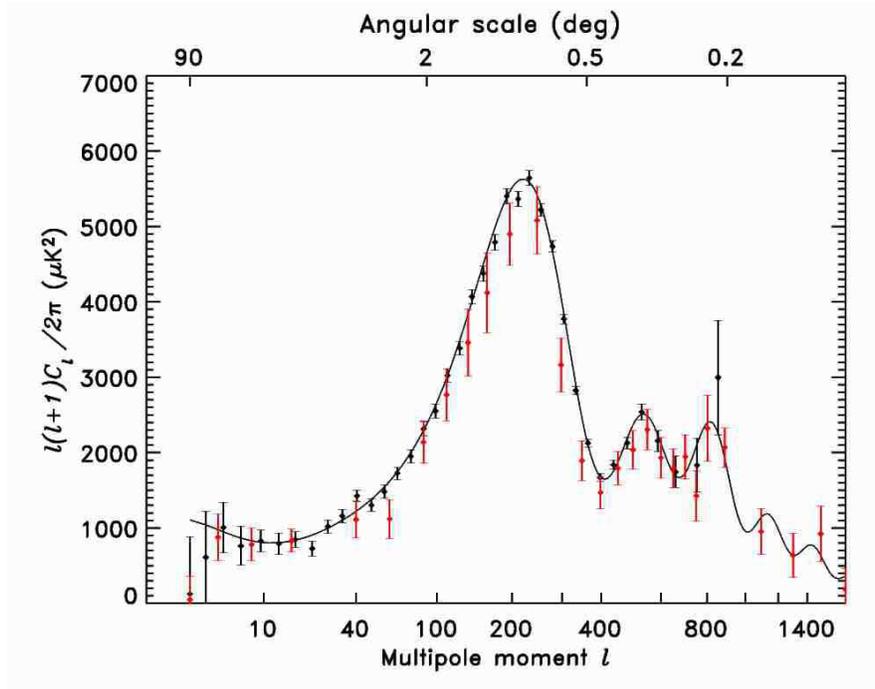}
\caption{
A compilation of data on fluctuations in the cosmic microwave background
radiation~\protect\cite{CMBexpts}. The darker (black) error bars are those 
from the WMAP satellite, the lighter (red) error bars are those of 
previous CMB 
experiments.} 
\label{fig:CMB}
\end{figure}

The next piece of evidence for the Big Bang to be discovered was the
cosmic microwave background (CMB) radiation~\cite{CMB}. Extrapolating the
present Hubble expansion back in time, the CMB is thought to have been
emitted when the Universe was about 3000 times smaller and hotter than it
is today, with age $\sim 3 \times 10^5$~y. The CMB has a dipole deviation
from isotropy at the $10^{-3}$ level, which this is believed to be due to
the Earth's motion relative to a Machian cosmological frame. Smaller-scale
anisotropies have been discovered more recently by the COBE satellite and
subsequent experiments, as seen in Fig.~\ref{fig:CMB}~\cite{CMBexpts}, and 
may provide a window on the Universe when it was much younger still, as we
shall see later.

\begin{figure}[htb]
\centering
\includegraphics*[width=85mm]{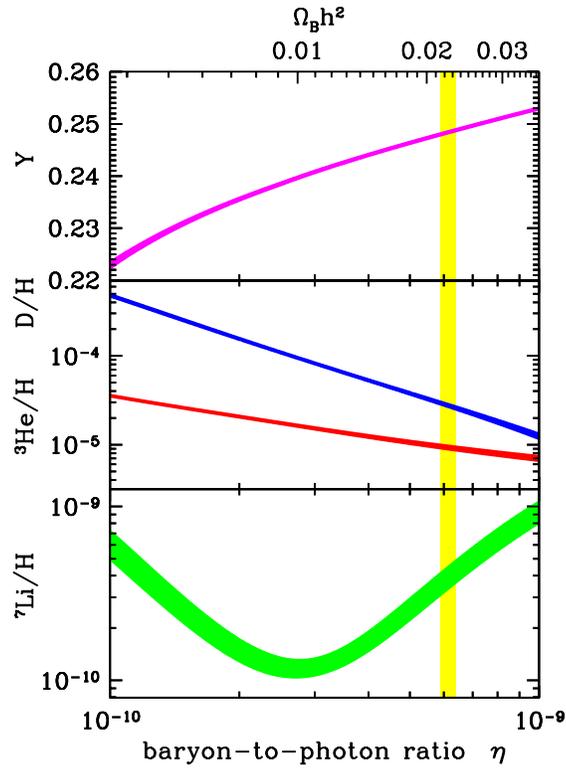}
\caption{
There is good concordance between the observed abundances of light
elements and calculations of Big-Bang nucleosynthesis~\protect\cite{BBN}.}
\label{fig:BBN}
\end{figure}

The third piece of evidence for the Big Bang was provided by the
abundances of light elements seen in Fig.~\ref{fig:BBN}, which are thought
to have been established when the Universe was about $10^8$ times smaller
and hotter than it is today~\cite{BBN}, with age $\sim 1$ {\rm to}
$10^2$~s. This nuclear `cooking' must have occurred when the temperature
$T$ of the Universe corresponded to characteristic particle energies $\sim
1$~MeV.

Back when the Universe was $\sim 10^{-6}$ to
$\sim 10^{-5}$~s old, it is thought to have made a transition from a
plasma of quarks and gluons to hadrons at a temperature $T \sim
100$~MeV~\cite{QHPT}. Previous to that, the electroweak transition when
Standard Model particles acquired their masses is thought have occurred
when the Universe was $\sim 10^{-12}$ to $\sim 10^{-10}$~s old, and the
temperature $T \sim 100$~GeV~\cite{EWPT}.

During the expansion of the Universe, it acts as a `cosmic decelerator', 
whose effective temperature $T$ falls as it expands~\cite{first3}:
\begin{equation}
T \; \sim \; {1 \over a},
\label{adiabatic}
\end{equation}
where $a$ is the scale factor measuring the size of the Universe. During 
the early history of the Universe when most particle masses were 
negligible, the rate of expansion was such that the age
\begin{equation}
t \; \sim \; a^2 \; \sim \; {1 \over T^2}.
\label{expansionrate}
\end{equation}
Inserting the units, one finds that the temperature of the Universe would 
have been about $10^{10}$~K when it was about a second old. Such high 
temperatures correspond to high energies for the thermalized particles: 
$10^{10}$~K $\sim 1$~MeV (cf the electron mass $\sim 1/2$~MeV), 
$10^{13}$~K $\sim 1$~GeV (cf the proton mass $\sim 1$~GeV). In general, 
the time-temperature relation is such that
\begin{equation}
t ( {\rm sec} ) \; \sim \; {1 \over T ( {\rm MeV} )^2}.
\label{timetemp}
\end{equation}
Thus, it is clear that the very early history of the Universe must have 
been dominated by elementary particles, and only their physics can explain 
how the Universe got to be the way it is today.

\section{Particle Physics beyond the Standard Model}

\begin{figure}[htb]
\centering
\includegraphics*[width=85mm]{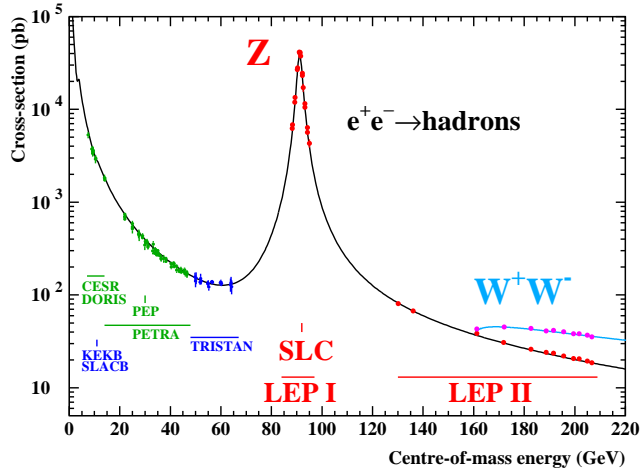}
\caption{
Data from LEP and other $e^+ e^-$ experiments agree perfectly
with the predictions of the Standard Model~\protect\cite{Grunewald}.}
\label{fig:LEP}
\end{figure}

As you can see in Fig.~\ref{fig:LEP}, data from the LEP accelerator are,
unfortunately, in excellent agreement with the Standard Model. Indeed, no
accelerator data provide evidence for physics beyond the Standard Model.
Nevertheless, particle physicists are convinced that there must be
accessible physics beyond the Standard Model, because it leaves many
fundamental questions unanswered. We seek the {\it Origin of Particle
Masses} and the reason why they are so much smaller than the Planck mass
$m_P \sim 10^{19}$ GeV. Are the masses due to a Higgs boson, and is it
accompanied by supersymmetric particles? We seek a {\it Theory of
Flavour}, because the Standard Model has six random-seeming quark masses,
three disparate charged-lepton masses, three weak mixing angles and the
CP-violating Kobayashi-Maskawa phase. Moreover, we seek a {\it Grand
Unified Theory}, because the Standard Model has three independent gauge
couplings and (potentially) a CP-violating phase in QCD. Altogether, the
Standard Model has a total of 19 parameters, without even addressing the
more fundamental questions of the origins of the particle quantum numbers.
Beyond all these beyonds, other theorists seek a {\it Theory of
Everything} that includes gravity, reconciles it with quantum mechanics,
explains the origin of space-time and why we live in four dimensions (if
we do so).

Non-accelerator neutrino experiments~\cite{SK,SNO} now provide us with the
first direct evidence for physics beyond the Standard Model, convincing us
that neutrinos oscillate and have different non-zero masses. To describe
these, we need three neutrino mass parameters, three neutrino mixing
angles and three CP-violating phases in the neutrino sector. Moreover, we
should not forget about gravity, with at least two parameters to
understand: Newton's constant $G_N \equiv m_P^{-2} \sim (10^{19}$
GeV)$^{-2}$ and the cosmological `constant', which recent data suggest is
non-zero~\cite{DM}, and may not even be constant. Talking of
cosmology, we would need at least one extra parameter to produce an
inflationary potential, and at least one other to generate the baryon
asymmetry, which cannot be explained within the Standard Model.

At what energy scale might new physics beyond the Standard Model appear,
between the energies $\lappeq 100$~GeV already explored and the Planck
energy $\simeq 10^{19}$~GeV? The Problem of Mass must presumably be solved
by new physics at some energy $\lappeq 1$~TeV, whether it be just a Higgs
boson or some richer physics such as supersymmetry\cite{susy}. Simple
ideas of Grand Unification suggest new physics at a scale $\simeq
10^{16}$~GeV~\cite{GUTs}. On the other hand, we have no good idea what
energy scale might be associated with the solution to the Problem of
Flavour, or where extra dimensions might appear. If there is a significant
discrepancy between the BNL measurement of the muon anomalous magnetic
moment~\cite{g-2} and the Standard Model, which is not yet established,
this could only removed by new physics at a scale $\lappeq 1$~TeV. there
are two circumstantial pieces of evidence in favour of Grand Unification,
namely the existence of neutrino masses - which might have been generated
at some mass scale between $\simeq 10^{10}$~GeV and $\simeq 10^{15}$~GeV -
and the weak neutral-current mixing angle $\sin^2 \theta_W$. The value of
the latter could be explained by Grand Unification at a scale $\simeq
10^{16}$~GeV combined with supersymmetry at a scale $\simeq
10^3$~GeV~\cite{GUTs}.

The LHC will be able to discover `any' new physics at a scale $\lappeq
10^3$~GeV, but many of the other ideas mentioned above may not be directly
testable at accelerators for the foreseeable future. Astrophysics and
cosmology may provide the only laboratories for testing some of these
ideas. For example, the cosmic microwave background (CMB) and inflation
may be providing a direct window on physics at the GUT scale, and
ultra-high-energy cosmic rays (UHECRs) might be due to the decays of
metastable superheavy particle weighing $\sim 10^{13}$~GeV or
so~\cite{cryptons}. On the other hand, many particle candidates for dark
matter weigh $\lappeq 1$~TeV and could be detected at the LHC. It seems
that accelerators, astrophysics and cosmology are condemned to symbiosis.

\section{Density Budget of the Universe}

What does the Universe contain? Let us enumerate its composition in terms
of the density budget of the Universe, measured relative to the critical
density: $\Omega_i \equiv \rho_i / \rho_{crit}$.

Inflation~\cite{inflation} suggests that the {\it total density} of the
Universe is very close to the critical value: $\Omega_{tot} \simeq 1 \pm
O(10^{-4})$, and this estimate is supported by CMB data~\cite{CMB}. I
remind you that inflation explains why the Universe is so large: the scale
size $a \gg \ell_P \sim 10^{-33}$~cm, why the Universe is so old: its age
$t \gg t_P \sim 10^{-43}$~s, why its geometry is so nearly flat with a
Euclidean geometry, and why the Universe is so homogeneous on large
scales.

It achieves these feats by postulating an epoch of (near-) exponential
expansion during the very early Universe, making the Universe very large
and giving it a long time to recollapse (if it ever will). Even the most
distant parts of the observable Universe would have been very close to
each other prior to this inflationary epoch, and so could have
synchronized their behaviours. This inflationary expansion would have
blown the Universe up like an inflated ballon, which seems almost flat to
an ant living on its surface. During the inflationary expansion, quantum
fluctuations in the inflaton field would have generated small density
perturbations (cf. the observations in Fig.~\ref{fig:CMB}) capable of 
growing into the structures seen in the Universe
today~\cite{infperts}, as discussed later.

Big-Bang Nucleosynthesis suggests that the {\it baryon density} $\Omega_b
\simeq 0.04$~\cite{BBN}, an estimate that has been supported by analyses
of the relative sizes of small fluctuations in the CMB at different
scales~\cite{CMB}.

The baryons are insufficient to explain the {\it total matter density}
$\Omega_m \simeq 0.3$, as estimated independently by analyses of clusters
of galaxies and, more recently, by combining the observations of
high-redshift supernovae with those of the CMB. The supernovae constrain
the density budget of the Universe in a way that is almost orthogonal to
the CMB constraint, and is very consistent with the prior indications from
galaxy clusters~\cite{DM}.

Observations of the structures that have formed at different scales in the
Universe suggest that most of the missing dark matter is in the form of
non-relativistic {\it cold dark matter}, as discussed in the
next session.

The theory of structure formation suggests that very little of the dark 
matter is in the form of {\it hot dark matter} particles that were 
relativistic when structures started to form: $\Omega_{hot} h^2 < 
0.0076$~\cite{CMBexpts}. Applying this constraint to neutrinos, for which
\begin{equation}
\Omega_\nu h^2 \; \simeq \; { \Sigma_i m_{\nu_i} \over 93 {\rm eV}},
\label{omeganu}
\end{equation}
this constraint tells us that $\Sigma_i m_{\nu_i} < 0.7$~eV, a limit that 
is highly competitive with direct limits~\cite{directnu}.

If $\Omega_{tot} \simeq 1$ and the matter density $\Omega_m \sim 0.3$, how
do we balance the density budget of the Universe? There must be {\it
vacuum energy} $\Lambda$ with $\Omega_\Lambda \sim 0.7$. All the available
cosmological data are consistent with $\Lambda$ having been constant at
redshifts $z \lappeq 1$, as per Einstein's original suggestion of a
cosmological constant. However, we cannot yet exclude some slowly varying
source of vacuum energy, `quintessence' with an equation of state
parametrized by $w \equiv p / \rho \lappeq - 0.8$~\cite{quint}. Measurable
vacuum energy would provide a second general-relativity observable to
explain, in addition to the Planck mass scale $m_P$. This would provide a
tremendous opportunity for any theory of everything including quantum
gravity, such as string. The ultimate challenge for theoretical physics
may be to calculate $\Lambda$.

\section{Cosmological Baryogenesis}

We have seen that Big-Bang Nucleosynthesis~\cite{BBN} and the
CMB~\cite{CMB} independently imply that baryons make up only a few \% of
the density of the Universe.  Numerically, this corresponds to a
baryon-to-photon ratio $n_b / n_\gamma \sim 10^{-9} - 10^{-10}$, raising
several questions. Why is there so little baryonic matter? Why is there
any at all? Why is there apparently no antimatter?

Astronauts did not disappear in a burst of radiation when they landed on
the Moon, and neither have space probes landing on Mars or an asteroid.
The small abundance of antiprotons in the cosmic rays is consistent with
their production by primary matter cosmic rays~\cite{BESS}, and no
antinuclei have been seen~\cite{AMS}. If there were any large
concentration of antimatter in our local cluster of galaxies, we would
have detected radiation from matter-antimatter annihilations at its
boundary. The CMB would have been distorted by similar radiation from any
matter-antimatter boundary within the observable Universe~\cite{deru}. So
it seems that there must be a real cosmological asymmetry betwen matter
and antimatter.

This could be explained if, going back to when the Universe was less than
$10^{-6}$~s old, it contained about one extra quark for every $10^9$
quark-antiquark pairs in the primordial soup. As the Universe expanded, 
most of the quarks would have annihilated with those antiquarks to produce 
radiation, and the few quarks left over would have survived to combine 
into the baryons seen today. Where did this small quark-antiquark 
asymmetry originate? Did the Big Bang start off with it, or did the laws 
of Nature generate it during the subsequent expansion?

The conditions for such cosmological baryogenesis were established by
Sakharov in 1967~\cite{BBB}. There has to be a difference between the
interactions of matter and antimatter particles, in the form of
charge-conjugation (C) violation, which was discovered in the weak
interactions in 1957, and CP violation, which was discovered in kaon
decays in 1964. There must also have been a departure from thermal
equilibrium, which would have been possible during a phase transition,
perhaps the electroweak phase transition when $t \sim 10^{-10}$~s or a GUT
phase transition when $t \sim 10^{-36}$~s, or at the end of inflation.
Finally, there must have been a violation of baryon number, which would
have happened through nonperturbative weak interactions at high
temperatures~\cite{tH} and is thought to be a generic feature of GUTs.

Various specific mechanisms for Big-Bang baryogenesis have been proposed,
ranging from the out-of-equilibrium decays of GUT bosons~\cite{Yosh} or
heavy neutrinos~\cite{FY} to processes around the epoch of the electroweak
phase transition. The CP violation in the Standard Model seems inadequate 
to generate the required baryon asymmetry, but this might be possible if
it is extended to include supersymmetry~\cite{LEBBB}.

\section{Formation of Structures}

How have these hard-won baryons organized themselves into the structures -
clusters, galaxies, stars, planets and us - that we see in the Universe
today? As already mentioned, the prime candidates for the seeds of these
structures are quantum fluctuations in the inflaton field, which would
have caused different parts of the Universe to expand differently and
generated a {\it Gaussian random field} of density 
perturbations~\cite{infperts}. If the
inflaton energy was roughly constant during inflation, these perturbations
would be {\it almost scale-invariant}, as postulated by astrophysicists.  
The CMB data shown in Fig.~\ref{fig:CMB} are consistent with both these 
properties. Accepting this
scenario, the magnitude of the primordial perturbations would be
related to the field energy density $\mu^4$ during inflation:
\begin{equation}
\left( {\delta T \over T } \right) \; \propto \; \left( {\delta \rho \over 
\rho } \right)  \; \propto \; \mu^2 G_N.
\label{perts}
\end{equation}
Inserting the magnitude of $\delta \rho / \rho \sim 10^{-5}$ oberved by 
the COBE and subsequent experiments~\cite{CMB}, one estimates
\begin{equation}
\mu \; \simeq \; 10^{16}~{\rm GeV},
\label{infscale}
\end{equation}
comparable with the GUT scale~\cite{GUTs}.

These primordial perturbations would have produced embryonic potential
wells into which the non-relativistic cold dark matter particles would
have fallen, while relativistic hot dark matter particles would have
escaped. In this way, cold matter particles would have amplified the
amplitudes of the primordial density perturbations, while the baryons were
still coupled to the relativistic radiation. Then, when the baryonic
matter and radiation `re-' combined to form atoms, they would have fallen
into the deeper potential wells prepared by the cold dark matter. This
theory of structure formation fits remarkably well the data on all scales
from over $10^3$~Mpc down to $\sim 1$~Mpc~\cite{CMB,2df}.

\section{Candidates for Dark Matter}

It is this agreement that provides the most stringent upper limit on the
possible hot dark matter such as neutrinos. As discussed earlier, most of
the dark matter is thought to be non-relativistic cold dark matter. There
are almost as many candidates for this as in the Californian gubernatorial
election but, as in that case, some of the candidates are more favoured
than others.

{\it Lightest Supersymmetric Particle}: The existence of supersymmetry at
relatively low energies $\lappeq 1$~TeV is motivated by the hierarchy
problem, namely why is the electroweak scale $m_W \ll m_P \sim
10^{19}$~GeV, the only candidate we have for a primary mass scale in
physics~\cite{hierarchy}. Alternatively, one may rephrase this question as
why the Fermi constant $G_F \gg G_N$, the Newton constant, or as why the
Coulomb potential $V_{Coulomb} \gg V_{Newton}$ in an atom. This can be
traced to the fact that $Z e^2 = O (1) \gg m^2 /m_P^2$, which is in turn
due to the fact that the masses of particles in atoms $m \la m_W \ll m_P$.

You might think it be sufficient to set $m_W \ll m_P$ by hand and forget
about the hierarchy problem. However, this is insufficient because
Standard Model loop corrections to the Higgs and/or $W$ mass are
quadratically divergent:
\begin{equation}
\delta m^2_{H, W} \; \simeq \; O \left( {\alpha \over \pi} \right) 
\Lambda^2,
\label{quadratic}
\end{equation}
which is $\gg m_W^2$ if the cutoff $\Lambda$ where the Standard Model 
breaks down and new physics appears $\sim m_{GUT}$ or $m_P$. These loop 
corrections can be controlled by postulating supersymmetry~\cite{susy}, 
which 
predicts that bosons and fermions appear in pairs with equal couplings. 
Since the divergences in boson loops are positive and those in fermion 
loops are negative, (\ref{quadratic}) is replaced by
\begin{equation}
\delta m^2_{H, W} \; \simeq \; O \left( {\alpha \over \pi} \right) \left( 
m_B^2 - m_F^2 \right),
\label{susy}
\end{equation}
which is $\lappeq m_W^2$ if
\begin{equation}
| m_B^2 - m_F^2 | \lappeq 1~{\rm TeV}^2.
\label{natural}
\end{equation}
Thus, the loop corrections to the electroweak scale may be made naturally 
small by postulating small differences between supersymmetric partner 
particles.

\begin{figure}
\begin{center}
\mbox{\epsfig{file=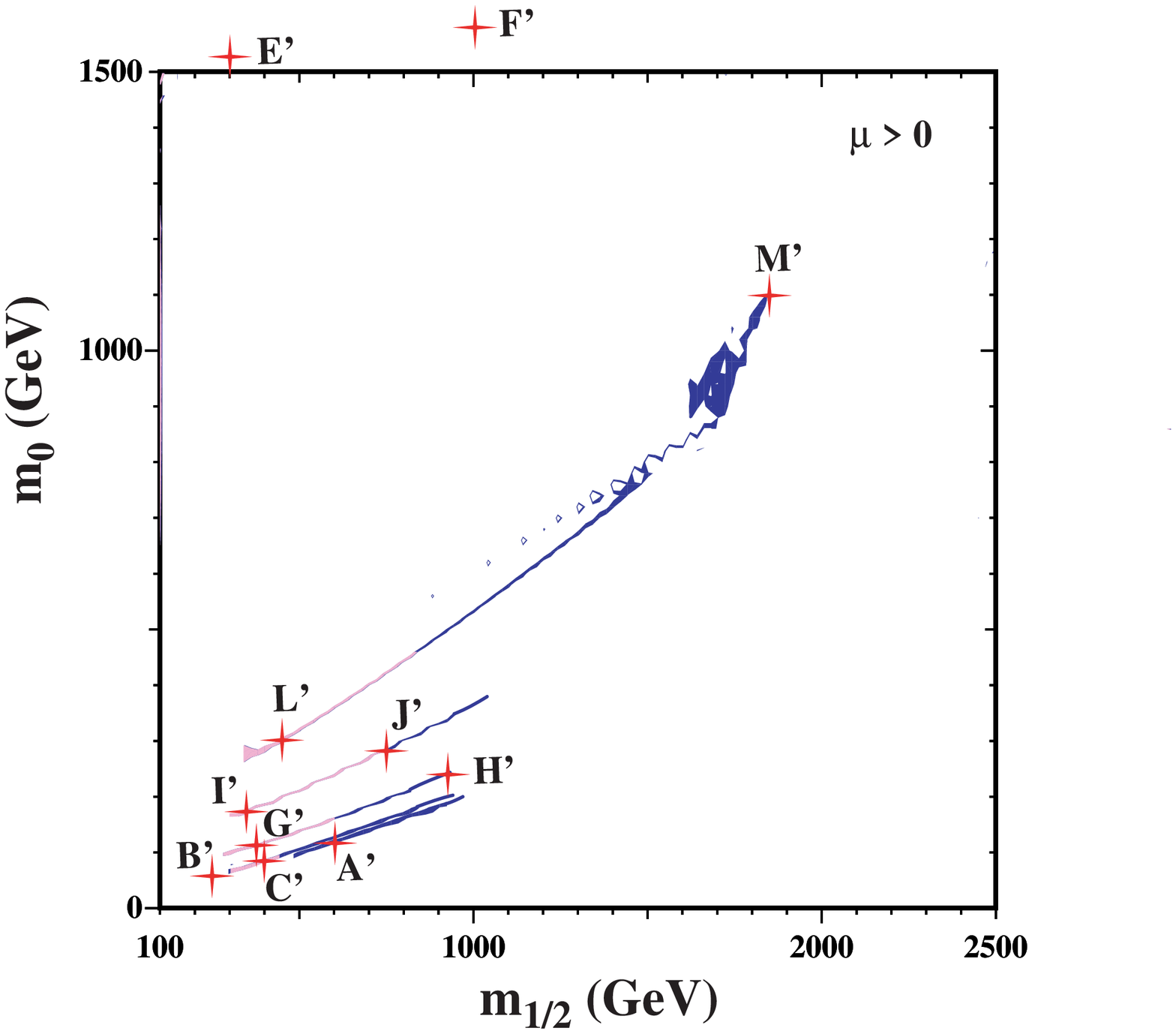,height=6cm}}
\mbox{\epsfig{file=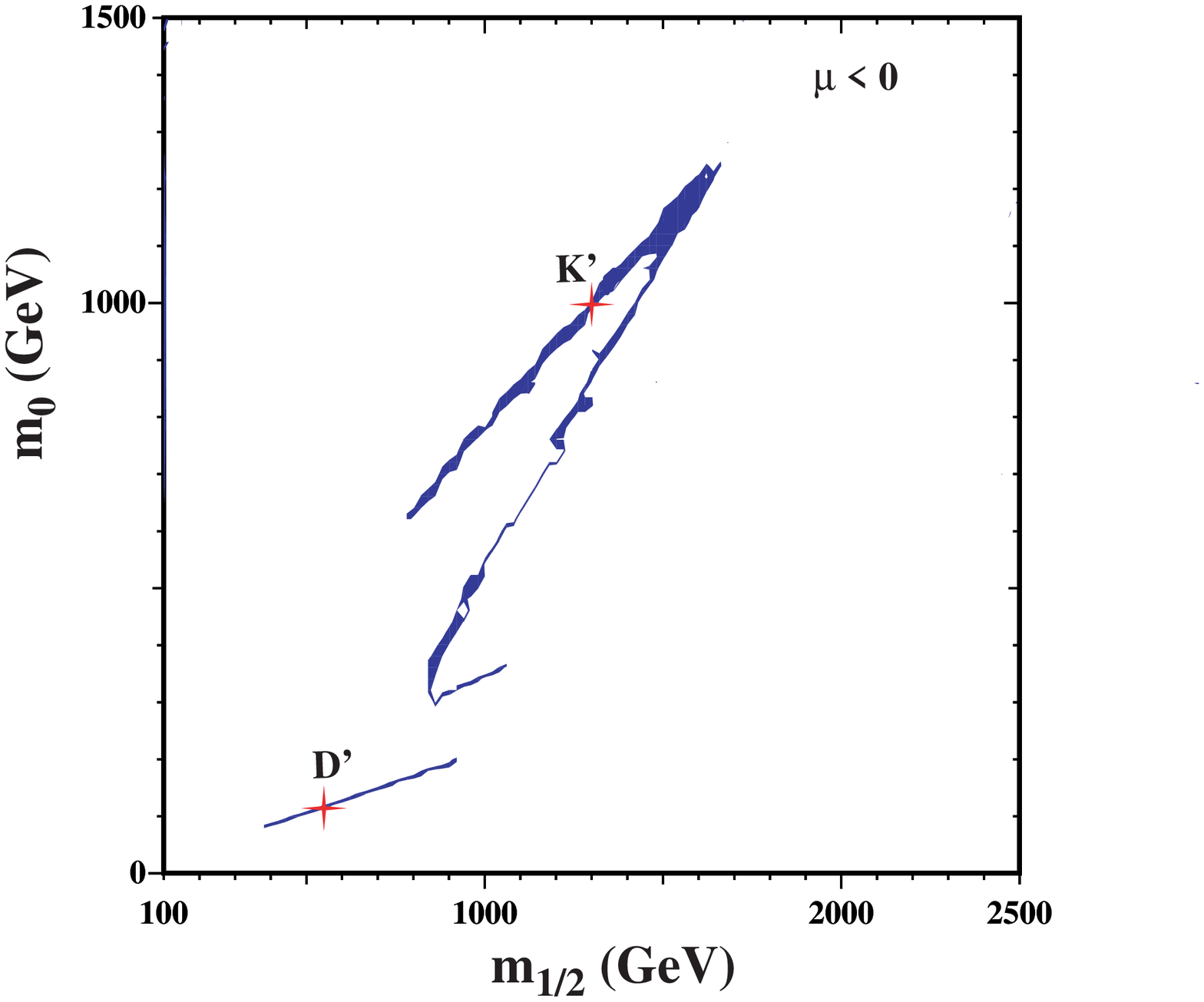,height=6cm}}
\end{center}
\caption{
Strips in the parameter space of the minimal supersymmetric extension of 
the Standard Model, assuming universal GUT-scale masses $(m_{1/2}, m_0)$ 
for the new supersymmetric fermions and bosons, respectively, that are 
consistent with accelerator and cosmological data~\cite{bench}. The 
different strips 
correspond to different values of the ratio $\tan \beta$ of Higgs vev's, 
and the two panels to different signs of the Higgs mixing parameter 
$\mu$. The crosses mark specific benchmark scenarios explored 
later~\cite{EFFMO}.} 
\label{fig:bench} 
\end{figure}

It is a generic feature of many supersymmetric models that the lightest
supersymmetric particle (LSP) is stable, as a result of a particular
combination of baryon and lepton number being conserved. This ensures that
heavier sparticles can only decay into lighter ones, and the LSP is stable
because it has no allowed decay modes. Furthermore, generically the LSP is
electrically neutral and has only weak interactions, making it an ideal
weakly-interacting massive particle (WIMP)~\cite{EHNOS}. Moreover, there 
are generic regions of the supersymmetric space where the relic LSP 
density falls within the range preferred by the cosmological data, as seen 
in Fig.~\ref{fig:bench}~\cite{bench}.

The LSP has many rivals to be the cold dark matter, including axions and 
the `Schwarzenegger' candidate, an ultraheavy metastable 
particle~\cite{Chung}. The 
next two sections discuss how these candidates might be elected by 
experiment.

\section{Searches for Dark Matter LSPs}

\begin{figure}[htb]
\centering
\includegraphics*[width=85mm]{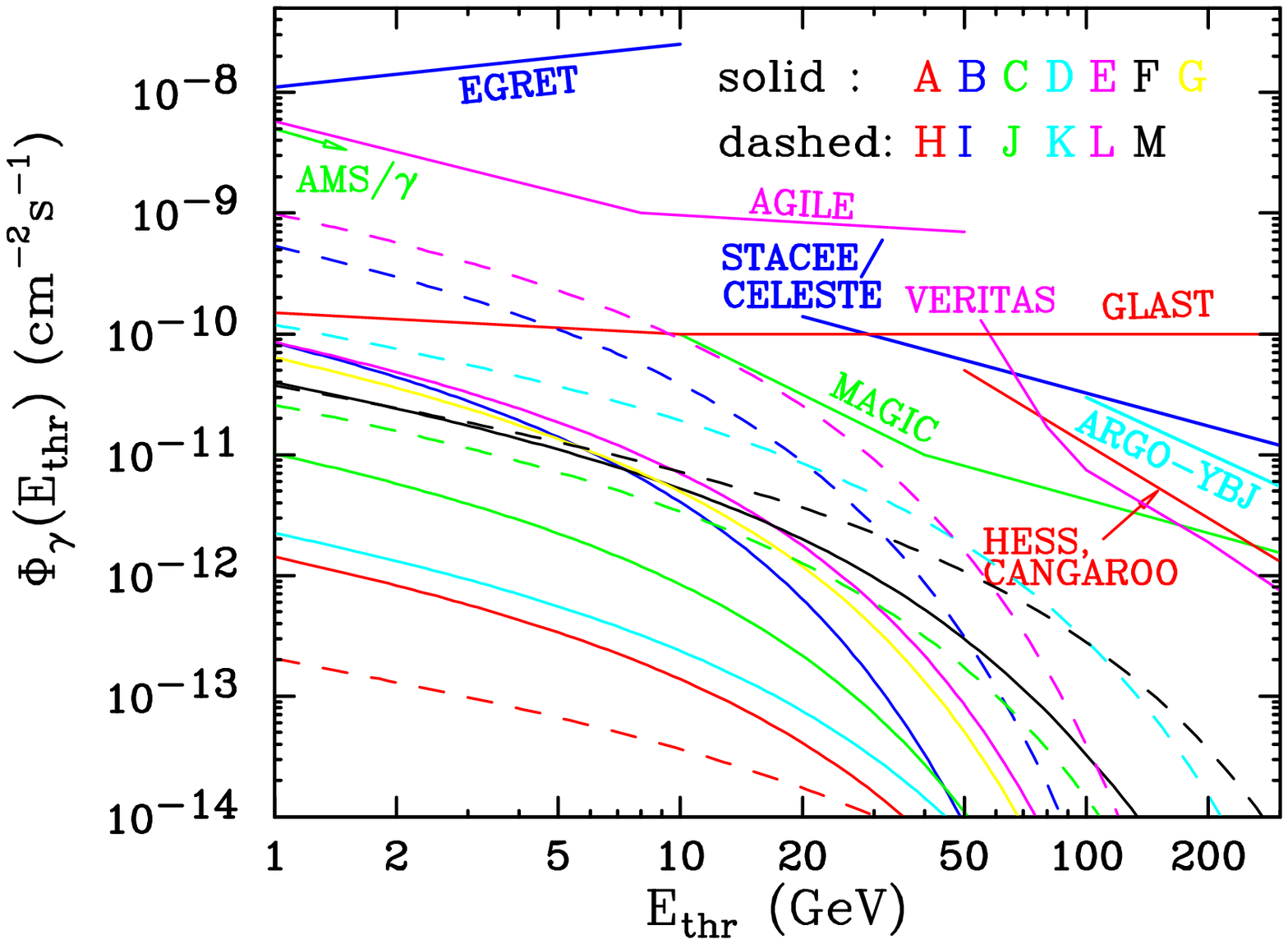}
\caption{
Observations of $\gamma$ rays from the galactic centre by GLAST and
ground-based experiments may be able to test certain
supersymmetric benchmark scenarios~\cite{EFFMO}.}
\label{fig:gamma}
\end{figure}

{\it Annihilations in the galactic halo}: These would produce some
antiprotons, positrons and photons that might be detectable among the
cosmic rays~\cite{halo}. As already discussed, the oberved antiprotons
appear completely consistent with production by primary matter cosmic
rays~\cite{BESS}. The prospects for detecting LSP annihilation positrons
do not look bright either, at least in a set of proposed supersymmetric
benchmark scenarios~\cite{EFFMO}. The prospects for detecting LSP
annihilation photons may be brighter, if the LSP density is enhanced in
the core of the galaxy. As seen in Fig.~\ref{fig:gamma}~\cite{EFFMO},
GLAST might be the best-placed to detect these.

\begin{figure}[htb]
\centering
\includegraphics*[width=85mm]{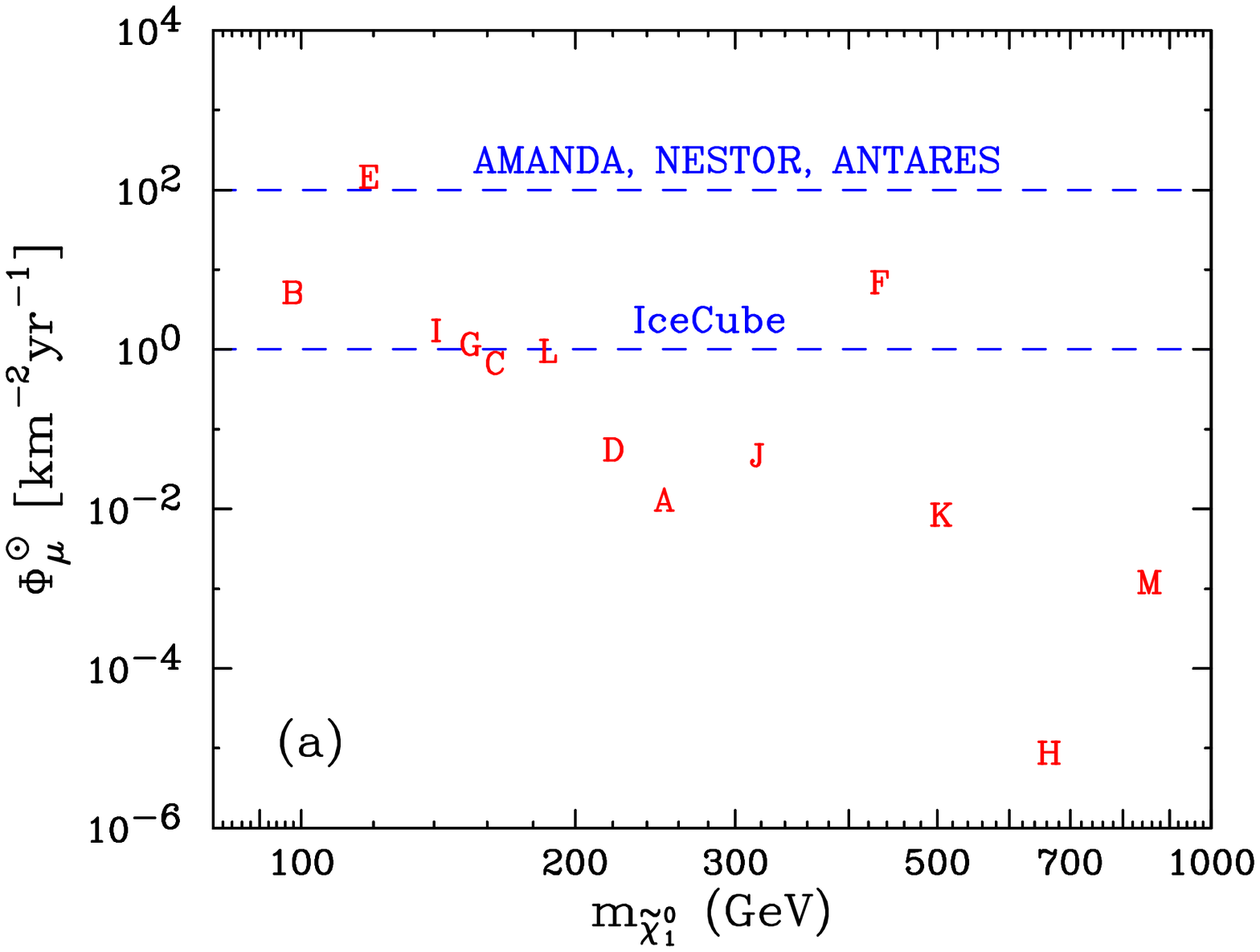}
\caption{
Searches in IceCube and other km$^2$ detectors for energetic muons 
originating from the interactions of high-energy
neutrinos produced by the annihilations of supersymmetric relic particles
captured inside the Sun may probe some supersymmetric benchmark
scenarios~\cite{EFFMO}.}
\label{fig:muons}
\end{figure}

{\it Annihilations in the Sun or Earth}: As LSPs fly though the galaxy,
some of them pass through the Solar System on hyperbolic trajectories. If
they pass through the Sun or Earth, they may scatter, and the deposit of
recoil energy may convert their trajectories into elliptical orbits with
perihelions (perigees) below the surface of the Sun (Earth). They would 
then
scatter repeatedly, losing more energy each time, until they eventually
settle into a cloud in the core. There they would annihilate, and any
high-energy neutrino they produce might be detectable in undergound
experiments~\cite{Sun}, either by interactions inside the detector or via
collisons in nearby material that produce muons passing through the
detector. As seen in Fig.~\ref{fig:muons}~\cite{EFFMO}, annihilations
inside the Sun would be detectable in several supersymmetric scenarios,
whereas the prospects for detecting terrestrial annihilations do not look
so good.

\begin{figure}[htb]
\centering
\includegraphics*[width=85mm]{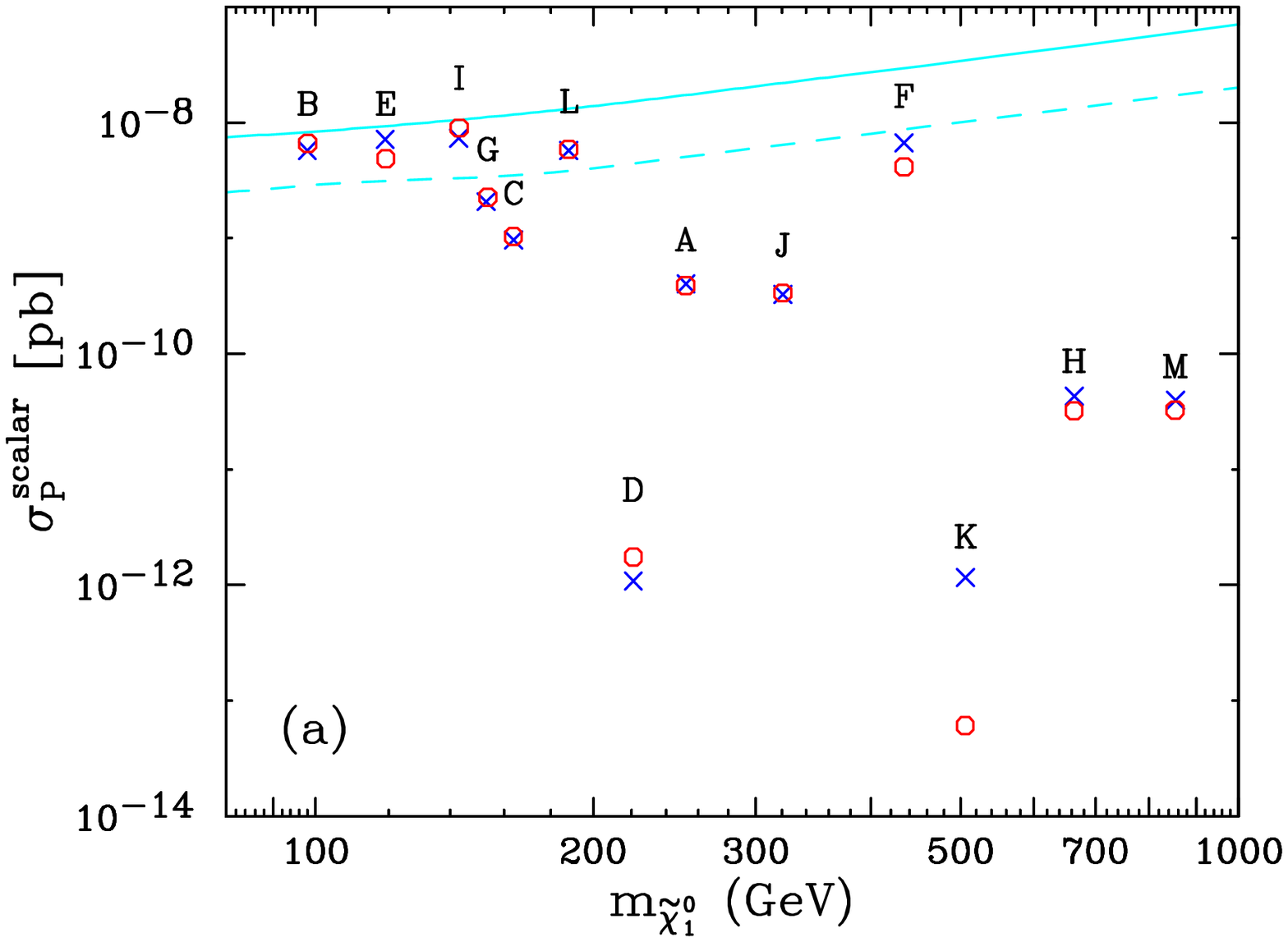}
\caption{
Direct searches for the scattering of superysmmetric relic
particles in underground detectors may probe some supersymmetric benchmark
scenarios~\cite{EFFMO}, compared with the possible sensitivities of 
future experiments.}
\label{fig:spinind}
\end{figure}

{\it Direct detection of dark matter scattering}: In many scenarios, it is
also possible to detect directly the scattering of LSPs on nuclei in a
low-background underground laboratory~\cite{GW}, via the few KeV of recoil
energy deposited. This scattering is expected to have both spin-dependent
and spin-independent components, with the latter seeming more promising
for the relatively heavy LSPs favoured by the absence of sparticles in
collider experiments to date. As seen in
Fig.~\ref{fig:spinind}~\cite{EFFMO}, dark matter may be detectable
directly in this way in a number of supersymmetric scenarios, at least in
some projected experiments.

\section{New Physics in Ultra-High-Energy Cosmic Rays?}

Now for the `Arnold' candidate~\cite{Chung}. The spectrum of cosmic rays 
falls almost
featurelessly $\sim E^{- \sim 3}$ until $E \sim {\rm few} \times
10^{19}$~GeV. At energies $E \gappeq 10^{20}$~GeV, protons or nuclei
coming from more than $\sim 50$~Mpc away would have scattered on CMB
photons before reaching us, producing pions and losing energy - the GZK
cutoff~\cite{GZK}.

The AGASA experiment~\cite{AGASA} has reported seeing ultra-high-energy
cosmic rays (UHECRs) beyond the GZK cutoff, but the HiRes
experiment~\cite{HiRes} does not. The Auger experiment~\cite{Auger} now
under construction in Argentina should be able to tell us definitively
whether such UHECRs exist. What might be their origins?

\begin{figure}[htb]
\centering
\includegraphics*[width=85mm]{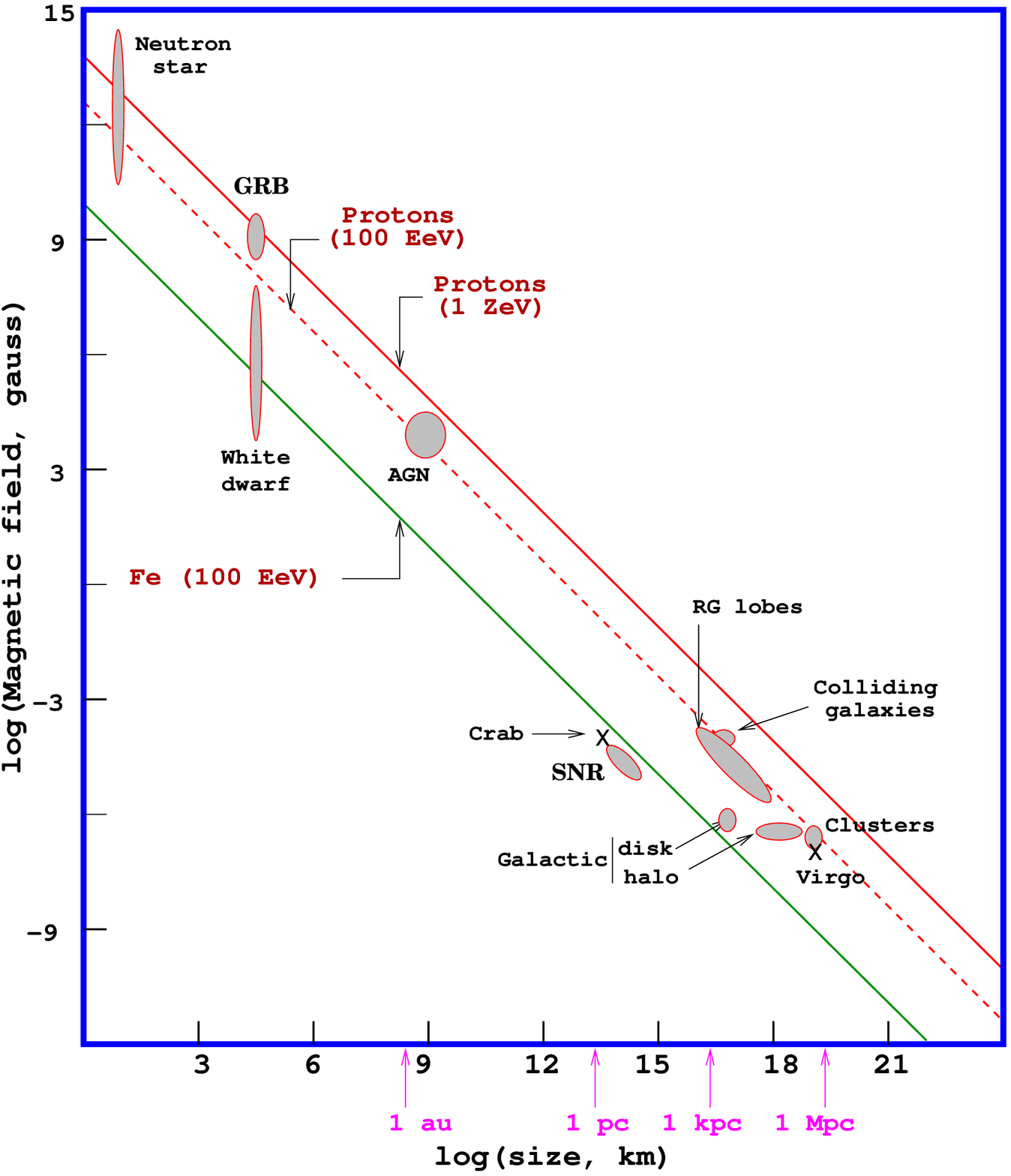}
\caption{
Schematic plot of the sizes and magnetic field strengths of possible 
astrophysical sources of UHECRs.}
\label{fig:Hillas}
\end{figure}

The most plausible is some `bottom-up' mechanism of acceleration by 
astrophysical sources. The upper limit on the energy attainable with such 
a cosmic accelerator is
\begin{equation}
E \; \sim \; 10^{18} Z \left( {R \over {\rm Kpc} } \right) \left( {B \over 
\mu G} \right)~{\rm eV},
\label{astroacc}
\end{equation}
where $Z$ is the atomic number of the accelerated nucleus, $R$ is the size
of the cosmic accelerator and $B$ its magnetic field.  
Fig.~\ref{fig:Hillas} shows some of the possible astrophysical sources of
UHECRs, including neutron stars, active galactic nuclei (AGNs),
radio-galaxy lobes and galactic clusters with gamma-ray bursters (GRBs)  
being other possible sources. In any scenario based on such discrete 
sources, one might expect clustering of arrival directions and 
correlations with astrophysical objects, as has sometimes been 
claimed~\cite{ST}.

\begin{figure}[htb]
\centering
\includegraphics*[width=85mm]{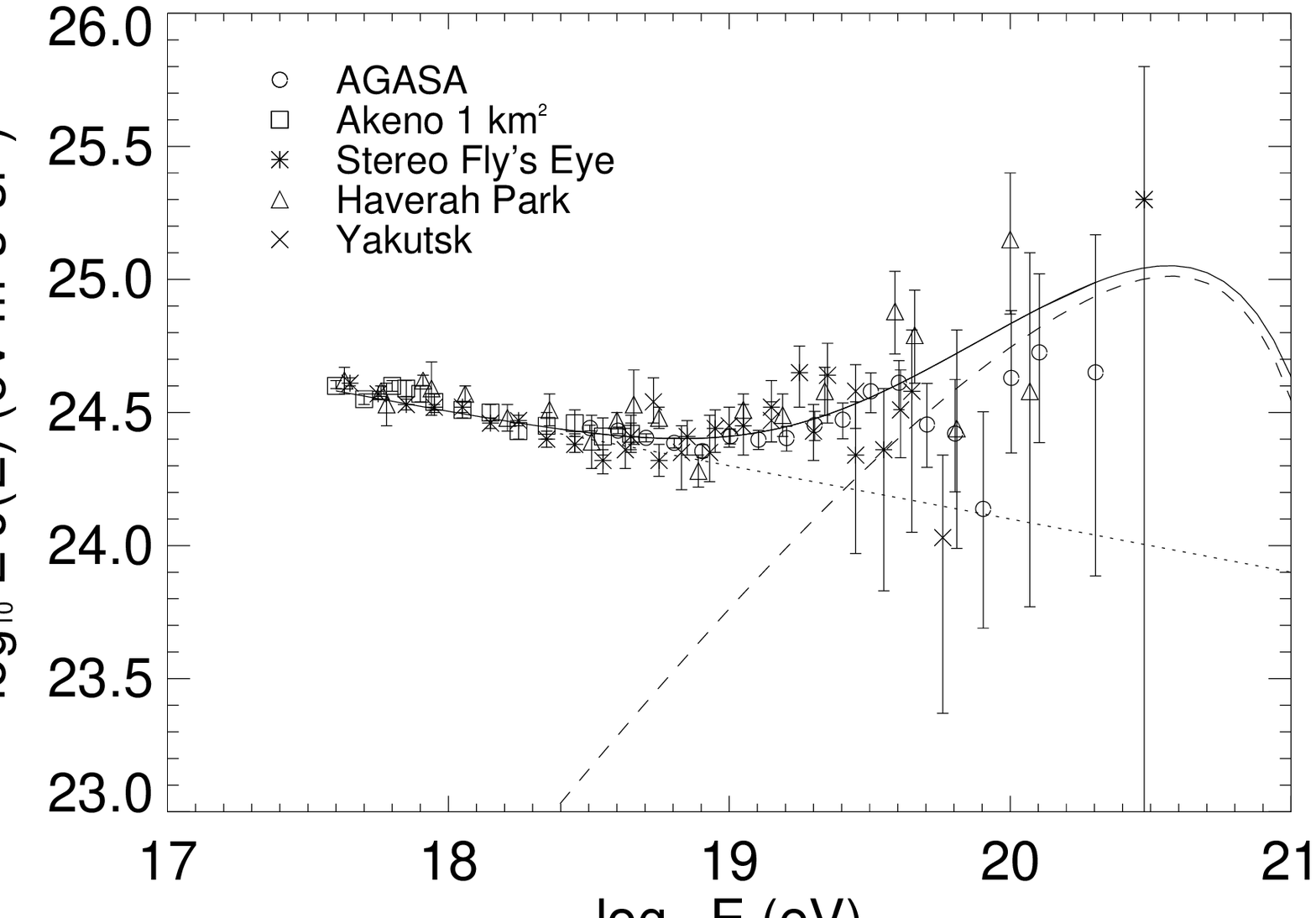}
\caption{
A fit to the UHECR spectrum including an extrapolation of the lower-energy 
spectrum (dotted line) and the decays of superheavy particles (dashed 
line)~\cite{SS}.}
\label{fig:Toldra}
\end{figure}

Alternatively, one might postulate some `top-down' model based on the
decays or interactions of massive GUT-scale particles~\cite{Berez}. Such
superheavy candidates include topological defects and metastable relic
particles, such as the cryptons expected as relics of the hidden sector in
string models~\cite{cryptons}. The energy spectrum of their decays can fit
the AGASA spectrum of UHECRs~\cite{SS}, as seen in 
Fig.~\ref{fig:Toldra}, but their composition is
potentially an issue. In such models, most of the UHECRs would arise from
decays within the halo of our own galaxy, and their arrival directions
should be anisotropic. There could be some clustering if a large fraction
of the cold dark matter in our galactic halo is clumped.

\section{The End of the Beginning}

Copernicus taught us that we do not live at the centre of the Universe.  
Modern astrophysicists teach us that we are not made of the same stuff as
most of the matter in the Universe, and modern cosmologists teach us that
matter is not even the dominant form of energy in the Universe. The
challenge for coming observations is to prove these assertions and
determine the nature of the missing matter and energy. The rest of this
summer institute will provide you with some of the tools you will need for
this task.

\end{document}